\documentclass[sigconf]{acmart}

\AtBeginDocument{%
  }

\copyrightyear{2025}
\acmYear{2025}
\setcopyright{rightsretained}
\acmConference[CHI EA '25]{Extended Abstracts of the CHI Conference on Human Factors in Computing Systems}{April 26-May 1, 2025}{Yokohama, Japan}
\acmBooktitle{Extended Abstracts of the CHI Conference on Human Factors in Computing Systems (CHI EA '25), April 26-May 1, 2025, Yokohama, Japan}\acmDOI{10.1145/3706599.3719708}
\acmISBN{979-8-4007-1395-8/2025/04}

\usepackage{hyperref}
\usepackage{enumitem}
\usepackage{subcaption}
\usepackage{algorithm}
\usepackage{algorithmic}
\usepackage{makecell}
\usepackage{colortbl}
\definecolor{green}{RGB}{185, 245, 183}
\definecolor{red}{RGB}{247, 167, 167}
\DeclareCaptionLabelFormat{subfigure}{(#2)}
\begin{document}
\newcommand{\Sys}{\textsc{GPTFootprint}}

\title[\Sys]{\Sys: Increasing Consumer Awareness of the Environmental Impacts of LLMs}


\author{Nora Graves}
\authornote{All three authors contributed equally to this research.}
\affiliation{%
  \institution{Princeton University}
  \city{Princeton}
  \state{New Jersey}
  \country{USA}
  \postcode{08544}
}
\email{eg5817@princeton.edu}

\author{Vitus Larrieu}
\authornotemark[1]
\affiliation{%
  \institution{Princeton University}
  \city{Princeton}
  \state{New Jersey}
  \country{USA}
  \postcode{08544}
}
\email{vl7131@princeton.edu}

\author{Y. Trace Zhang}
\authornotemark[1]
\affiliation{%
  \institution{Princeton University}
  \city{Princeton}
  \state{New Jersey}
  \country{USA}
  \postcode{08544}
}
\email{yingyue@princeton.edu}

\author{Joanne Peng}
\affiliation{%
  \institution{Princeton University}
  \city{Princeton}  
  \state{New Jersey}
  \country{USA}
  \postcode{08544}
}
\email{jzp@princeton.edu}

\author{Varun Nagaraj Rao}
\affiliation{%
  \institution{Princeton University}
  \city{Princeton}  
  \state{New Jersey}
  \country{USA}
  \postcode{08544}
}
\email{varunrao@princeton.edu}

\author{Yuhan Liu}
\affiliation{%
  \institution{Princeton University}
  \city{Princeton}  
  \state{New Jersey}
  \country{USA}
  \postcode{08544}
}
\email{yuhanl@princeton.edu}

\author{Andrés Monroy-Hernández}
\affiliation{%
  \institution{Princeton University}
  \city{Princeton}  
  \state{New Jersey}
  \country{USA}
  \postcode{08544}
}
\email{andresmh@princeton.edu}



\begin{abstract}

With the growth of AI, researchers are studying how to mitigate its environmental impact, primarily by proposing policy changes and increasing awareness among developers. However, research on AI end users is limited. Therefore, we introduce \Sys, a browser extension that aims to increase consumer awareness of the significant water and energy consumption of LLMs, and reduce unnecessary LLM usage. \Sys\  displays a dynamically updating visualization of the resources individual users consume through their ChatGPT queries. After a user reaches a set query limit, a popup prompts them to take a break from ChatGPT. In a week-long user study, we found that \Sys\  increases people's awareness of environmental impact, but has limited success in decreasing ChatGPT usage. This research demonstrates the potential for individual-level interventions to contribute to the broader goal of sustainable AI usage, and provides insights into the effectiveness of awareness-based behavior modification strategies in the context of LLMs.
\end{abstract}




\begin{teaserfigure}
\centering
\includegraphics[width=0.99\textwidth]{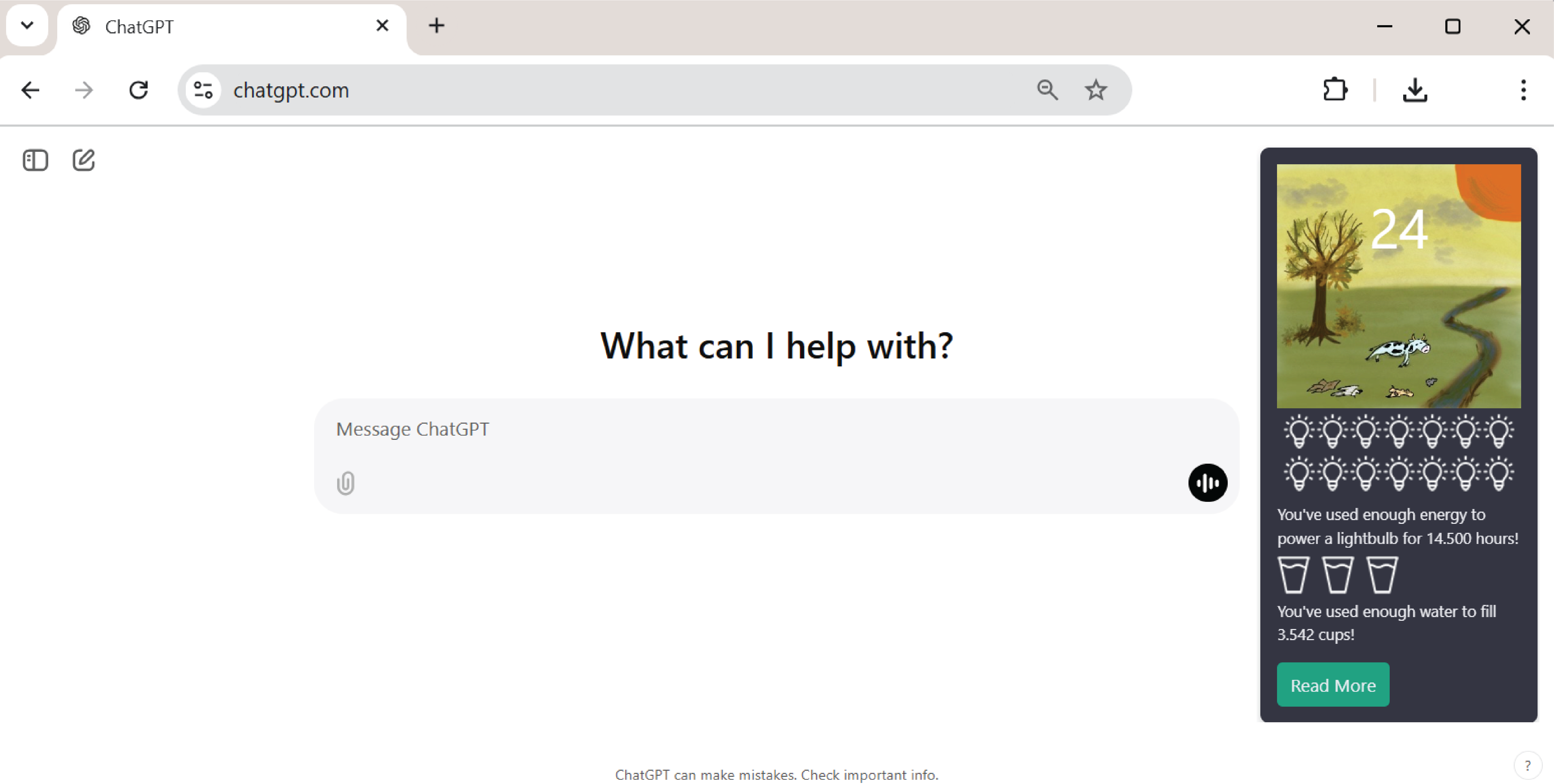}
  \Description[A screenshot of \Sys\ in use.]{This image displays a sample ChatGPT interface. On the right side of the screen is a side panel showing an illustrated image of a cow lying on the ground with lots of trash, a polluted sky, a dead tree, and a dried-up stream. Overlaid above that image is an Eco Score of 24. The text in the sidebar says ``You've used enough energy to power a lightbulb for 14.500 hours! You've used enough water to fill 3.542 cups!'' There is a pictogram of 14 lightbulbs, and another pictogram of 3 cups of water.}
  \caption{\Sys\ displays the amount of energy and water consumed in user-friendly terms and with pictogram icons (hours powering a light bulb and cups of water) in a side panel on top of the ChatGPT interface. In addition, there is an Eco Score, which can range from 0 to 100.}
  \label{fig:teaser}
\end{teaserfigure}


\maketitle

\section{Introduction}

The rapid advancement and widespread adoption of large language models (LLMs) has transformed how the public interacts with artificial intelligence. Tools like ChatGPT handle millions of queries daily \cite{chatstat}, assisting with tasks ranging from coding \cite{Codegen} to creative writing \cite{Confed}. However, this convenience comes at significant environmental costs that remain largely invisible to the end users. A single trained LLM has the carbon footprint equivalent to hundreds of households' annual emissions \cite{UnevenDistofAI}, processing a single query consumes ten times more energy than a standard web search \cite{Iea}, and cooling data centers requires substantial water consumption \cite{UncoveringAndAdressing}. As LLMs become increasingly integrated into daily workflows, establishing sustainable usage patterns early could significantly impact their long-term environmental footprint. Current solutions for quantifying LLM environmental impacts focus on industry-level interventions or expert-oriented tools \cite{QuantifyingCarbonEmissionsML, CodeCarbon}, leaving a gap for consumer awareness and engagement. While several tools track query counts \cite{ChatterClock, QuestionCount}, none provide accessible environmental impact metrics to end users.

In this paper, we present \Sys, a novel browser extension that addresses this gap by displaying dynamic environmental impact metrics during ChatGPT usage. Our approach innovates in three key ways: dynamic feedback that converts technical measures into human-scale units, privacy-preserving tracking that does not access query content, and a visually engaging interface that makes a user's impact simple to understand at a glance.

This research addresses three questions:

\begin{enumerate}
    \item How can we effectively communicate the environmental impact of individual LLM usage to end users?
    \item What metrics and visualizations most effectively help users understand their environmental impact?
    \item To what extent does dynamic environmental impact feedback influence people's use of LLMs?
\end{enumerate}

We evaluated \Sys\ with 9 participants who used it for 7 days. Participants reported a greater awareness of the environmental impact of LLMs, and expressed appreciation for the new understanding they gained. Participants also agreed that human-scale metrics and visualizations significantly improved understanding of their impact. Although awareness increased, behavioral change was limited by the utility of ChatGPT and sentiment about the limits of personal responsibility. These findings contribute to both environmental computing and human-computer interaction fields by demonstrating effective strategies for communicating AI environmental impact to end users and identifying barriers to behavior change. 
\section{Related Work}

\subsection{Environmental Impacts of LLMS}

In the past few years, research on the flaws and dangers of LLMs has grown, warning of broad societal and environmental impacts \cite{DangersofStochasticParrots}. Although certain machine learning applications can help mitigate climate change \cite{AligningAI, TacklingClimateChange, ClimateDoesNotExist}, they often also exacerbate climate change through emissions-intensive training and usage \cite{RisingCostsofTraining,DangersofStochasticParrots,UncoveringAndAdressing}. Current proposed solutions \cite{AligningAI, EnergyPolicyConsid, DangersofStochasticParrots,BringingEnergyEff} focus on governmental policies or industry-level energy conservation techniques instead of user behavior changes. Some researchers have attempted to engage individuals' awareness of carbon emissions from large scale computing \cite{QuantifyingCarbonEmissionsML, CodeCarbon}, but these programs require detailed knowledge of the hardware, hours used, compute providers, geographical regions, and more. Furthermore, they exclude other crucial environmental impacts like the water data centers consume \cite{UncoveringAndAdressing}. Although several ChatGPT extensions can count queries \cite{ChatterClock, QuestionCount}, none of these display an estimate of the resulting environmental impact. We saw a need for consumer access to dynamic, updating statistics about the environmental impact of their personal LLM usage. 

\subsection{Technology for Behavior Change}
Habit adjustment technology aims to increase positive habits and decrease negative ones. Awareness alone can influence actions, as with behavior-informing technology that display health statistics like step count in order to increase activity levels \cite{pedometer, TheoryDrivenDesign}. Because studies of physical fitness focus on promoting positive behaviors, like exercise, rather than reducing negative ones, like excessive LLM usage, we also draw from research on screen time reduction, which relies on strategies like gamification and goal setting that can effectively reduce both device and app-specific usage over time \cite{EvalEffAppsDes, PersontalTrackingScreen}. However, complete app lock-outs can cause frustration, which sometimes results in an ultimate failure to reduce screen time \cite{GoalKeeper, LockoutPhones}. It may be more effective to allow users to continue using the software, but make it increasingly challenging for them to do so \cite{InputManipulation}. 

\subsection{Eco-Feedback Technology}

Eco-feedback systems often incorporate user awareness by presenting information about personal environmental impact \cite{EcoFeedback}, which can increase environmentally friendly behaviors in transportation \cite{EcoEffPref, UbiGreenTransport}, water usage \cite{WaterBot, WaterFeedback}, and more \cite{EcoFeedback}. Eco-feedback systems that contextualize energy consumption perform better than those that do not \cite{CO2toKWh}. Although eco-feedback has been successful in various applications, they can lack significant long-term effects \cite{BeyondEnergyFeedback}. Researchers have suggested that speculative design \cite{BeyondEnergyFeedback} and social comparison components \cite{SocialSmartMeter} can be potential solutions for the lack of long-term effects.

\section{\Sys}

\begin{figure*}[!htbp]
    \centering
    
    \begin{subfigure}[b]{0.3\textwidth}
        \centering
        \includegraphics[width=\textwidth]{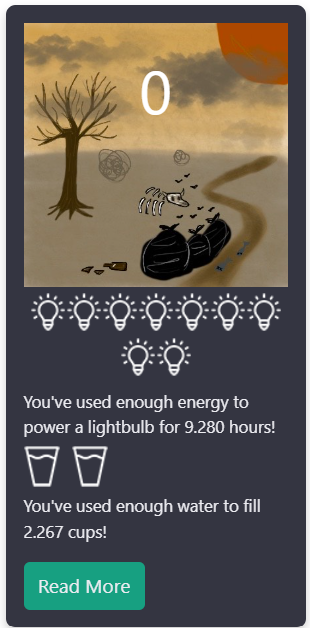}
        \caption{Side Panel}
        \label{fig:system design side panel display}
        \Description[Side Panel]{The side panel includes an illustration of a cow carcass, a dead tree, a trash bag, a dried-up stream, and a polluted sky. Overlaid on top is an Eco Score of 0. The text reads: You've used enough energy to power a lightbulb for 9.280 hours! You've used enough water to fill 2.267 cups! There is a pictogram showing 9 lightbulbs, and another pictogram showing two cups of water. At the bottom is a green button labeled Read More!}
    \end{subfigure}
    \begin{subfigure}[b]{0.6\textwidth}
        \centering
        \includegraphics[width=\textwidth]{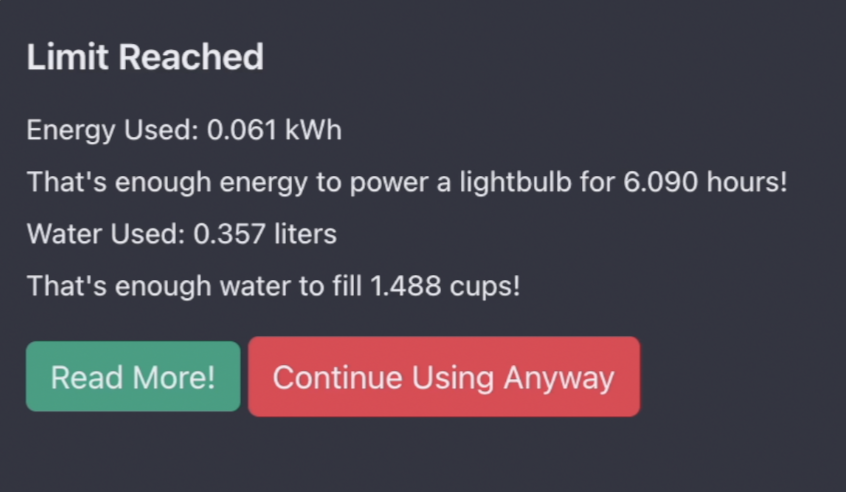}
        \caption{Popup}
        \label{fig:system design popup}
        \Description[Popup]{The limit popup displayed here includes the following text. Limit Reached. Energy used: 0.061 kWh. That's enough energy to power a lightbulb for 6.090 hours! Water used: 0.357 liters. That's enough water to fill 1.488 cups! At the bottom of the popup is a green button labeled Read More! and a red button labeled Continue Using Anyway.}
    \end{subfigure}
    \caption{System Display. (a) The side panel is always visible on screen, and displays an updating display of energy and water usage, with a link to read more about the environmental impacts of LLMs. (b) The popup appears each time the user consumes a certain amount of energy or water. Users can choose to read more, or close the popup and continue using ChatGPT.}
    \label{fig:intervention1}
\end{figure*}

We implemented \Sys\ as a Chrome Extension that layers on top of the ChatGPT website for ease of use. It consists of two main features: an Environmental Impact side panel (Figure \ref{fig:system design side panel display}) and a Limit Reached popup (Figure \ref{fig:system design popup}). Both link to the same Read More document, which provides further information about the environmental impact of LLMs. During user studies, the extension also included a server component, which saved information about ChatGPT and Extension usage for each participant. No version of \Sys\ accesses or stores the contents of queries or responses. All components were developed using JavaScript and CSS.

\subsection{Unit Conversion Techniques}
\Sys\ calculates energy and water consumption based on an average per-query value of 2.9Wh of energy \cite{Iea} and 16.9mL of water \cite{UncoveringAndAdressing}. Although the size of queries and responses affects energy and water consumption \cite{CodeCarbon, QuantifyingCarbonEmissionsML}, determining these variables would require accessing user queries. To mitigate security and privacy concerns, we chose to use per-query average values instead. Similar per-query average approaches to evaluating carbon emission costs from LLMs are standard for assessing model impacts \cite{estimatingcarbonfootprintbloom, Goldman_Sachs, semianalysis2023}, and even research that distinguishes between various query types finds that text generation and summarization, the two types relevant to \Sys, have a similar per-query average energy consumption \cite{huggingface}.

\subsection{Eco Score}
Pilot studies on a previous implementation of \Sys\ without Eco Score suggested that although participants appreciated seeing usage metrics, they typically did not enjoy using \Sys, due to feelings like guilt and stress, and it elicited little to no behavioral change. Therefore, we decided to implement Eco Score as a form of gamification, which can affect behavior \cite{EvalEffAppsDes} by inspiring intrinsic motivation \cite{GamificationIntrinsic, GamificationIntrinsicEducation} and increasing positive emotions like achievement \cite{Gamification}. We based the Eco Score on U.S. university grading systems, with which all participants would be familiar (in this system, 90-100 is excellent, 80-89 is good, and lower scores are average, poor, or a failure \cite{USGrading}). Such grading systems can induce motivation, due to the subconscious association with schooling \cite{classroomGrading}. Other sustainability measurements like Energy Star Scores \cite{energystar_score} and household carbon calculators \cite{CodeCarbon, carbontrust_sme_calculator} use similar systems. We designed the scoring algorithm with several goals in mind, as follows (the full algorithm appears in Appendix \ref{asec:algo}).

First, an average user should see an Eco Score low enough to encourage behavior change, but not low enough to cause frustration, which can result in failure to change behavior \cite{GoalKeeper, LockoutPhones}. Difficult but achievable goals lead to more motivation than easy or impossible goals \cite{GamificationIntrinsic}, so we aimed to place an average user just below a `good' score. In our pilot studies, participants queried 6 times per day on average. If these queries are each an hour apart, they will reach a score of 76, just below a `good' score. 

Second, the Eco Score should encourage efficient ChatGPT usage. We quantify efficient usage by the pauses in between queries, where longer pauses suggest that participants seek other resources when ChatGPT is unhelpful, and shorter pauses suggest that users rely solely on ChatGPT, even for tasks it may not be suited for. Therefore, queries in quick succession have a greater negative impact on the Eco Score than queries with longer pauses in between. Thus, all queries more than an hour apart lose 7 points each and all queries less than a minute apart lose 13 points, with several tiers in between. 

Third, an efficient day of ChatGPT usage should have little to no effect on the Eco Score the next day, while a particularly inefficient day of ChatGPT querying should have a noticeable impact the following day. We accomplished this with our score increase rate of 1 point every 20 minutes. Assuming an 8 hour night \cite{humanSleep}, the Eco Score will increase by 24 points overnight. Therefore, an average efficient user, with 6 queries each an hour apart, as described above, will start the next day with a score of 100 again, while a user with a lower Eco Score will not.

\subsection{Chrome Extension Design}
As soon as a user opens the ChatGPT website \footnote{Website at https://chatgpt.com/}, the extension requests current usage statistics from local Chrome storage and creates the Environmental Impact side panel. \Sys\ uses status codes as a proxy for queries: the extension listens for a successful POST request (statusCode === 200) made to the ChatGPT API (https://chatgpt.com/backend-api/conversation), ignoring requests containing `init' or `implicit', which represent tasks other than queries. Each query detection updates the locally stored query count, side panel, and popup. During the user studies, the extension also externally saved information about each participant's usage in a private Google Sheets spreadsheet. Each time the extension detected an event, such as query or popup opening, it logged the user ID, the date and time, and the event type (i.e. query, popup\_opening, popup\_closed, readmore\_clicked). This data is de-identified with a random user ID, and Google Chrome encrypted all information transmitted between users and the server.
\subsubsection{Environmental Impact Side Panel}
The side panel (Figure \ref{fig:system design side panel display}) initially appears on the top right of the screen, but can be dragged to another area to avoid obscuring ChatGPT's interface. The side panel is always present, so the user cannot minimize it or drag it offscreen. With each query, the side panel updates its display to reflect the total energy and water used in two display types. First, the Eco Score graphic provides users with a glanceable metric of their environmental impact. The score is displayed on top of an image, which changes dynamically with the score, depicting a more polluted, bleak environment when the score drops and a cleaner, brighter environment when the score increases. Second, the display of human-scale metrics and corresponding pictogram charts \cite{lightbulb_icon, cybertruck_icon, water_glass_icon, bath_tub_icon, hot_tub_icon} contextualize impact in glanceable and easily understandable terms for a typical end user of ChatGPT. The energy metric dynamically shifts from hours powering a light bulb to miles driving a Tesla as energy consumption increases, while the water metric shifts from cups to bathtubs to hot tubs. At the bottom of the side panel is a button labeled Read More, which links to information about LLMs and their environmental impact.
\subsubsection{Limit Reached Popup}
The limit popup (Figure \ref{fig:system design popup}) automatically appears each time the user reaches a certain energy or water limit. It cannot be moved, but will disappear once the user clicks the ``Continue Using Anyway'' button. The popup displays the total energy and water consumed in both user-friendly metrics and standard metrics (kWh and liters), and it includes the same Read More button. Originally, the limit was three queries, but pilot participants thought this was too frequent. We increased the limit to seven queries, which was successful during a second pilot study. 

\section{Methods}
We evaluated \Sys\ with a week-long user study with nine participants, all college students between ages 18-24. We recruited current ChatGPT users from university courses that allow LLM usage, and through snowball sampling. The study received IRB approval. We informed all subjects of the study and they each consented to our procedures before participating.

First, we sent users a pre-survey (Appendix \ref{apdx: pre survey}), asking about their current concern for the environment, along with a link and instructions for downloading and setting up \Sys. For the next week, we asked them to only use ChatGPT in Chrome with the extension enabled. After the trial period ended, we sent participants an exit survey (Appendix \ref{apdx: post survey}), in which they answered multiple-choice questions (using the Likert scale from 1 to 5) and wrote open-ended responses about whether they enjoyed \Sys, how it made them feel, and whether it impacted their behavior. We coded each response, then grouped codes into five themes (Appendix \ref{apdx: codebook}), finding key quotes for each theme. In addition, the exit survey contained instructions for determining ChatGPT usage before and during the trial using a Colab Notebook (Appendix \ref{apdx: colab notebook}). Participants uploaded their ChatGPT history locally to the Notebook, then reported the outputted query counts to our exit survey. Thus, we systematically collected usage statistics from before and after the study period without accessing participant's private conversation records. 

\section{Results}
We evaluated \Sys\ on its ability to increase user awareness of environmental impact and to decrease ChatGPT usage. In our analysis, three themes emerged: the importance of personalized information, the emotional impact of \Sys, and the difficulty of decreasing the usage of a valuable tool like ChatGPT.

\subsection{Participants Appreciate New Awareness of Their Personal Environmental Impact}

Participants uniformly agreed that \Sys\ was very educational, with an average response of 4.11 (SD = 0.782, median = 4) on a Likert Scale ranging from 1 (learned nothing from \Sys) to 5 (learned a lot). All participants noted a new awareness of their impact, which led many to some personal reflection: \textit{``This really put my usage into perspective and made me think about whether the use of ChatGPT was worth its cost or not''} [P4], \textit{``Frankly, it was more water than I drink in a regular basis, so it was eye opening''} [P2]. Additionally, almost half of the participants (4 out of 9) reported caring more about the impact of their ChatGPTs after the study than before, while the rest reported no change; the average score increased from 3.0 out of 5 (SD = 1, median = 3) before the trial to 3.55 (SD = 0.882, median = 4) afterwards. 

In pilot studies, participants appreciated the real-world, contextualized metrics, but found the side panel wordy, and therefore confusing, with many requests for more visuals. In this iteration of \Sys, therefore, we condensed the text and added pictograms. These updates were successful. No participants reported confusion about the metrics, and many praised their real-world applicability: \textit{``things I was familiar with (cups of water and hours for a lightbulb), and not just random numbers...had a more significant effect''} [P3]. The visuals, too, impacted participants: \textit{``having the visual indicator for the cost of my queries would reduce my usage''} [P8].

Notably, participants discussed not only their new awareness itself, but also their appreciation for that awareness. In fact, three different participants wrote about how seeing \textit{``reminders''} of ChatGPT's environmental impact is \textit{``important''} [P2, P5, P7]. Most users were likely to continue using \Sys\ after the trial ended (often citing the importance of awareness as a reason), rating the likelihood of future use 3.44 out of 5 on average, with only one participant responding with a value below 3. This desire to understand the hidden environmental impacts of ChatGPT emphasizes the importance of \Sys, and similar programs, as LLMs like ChatGPT continue to grow in popularity.

\subsection{\Sys\ Elicits Strong Emotional Responses}
Another theme among participants was that \Sys\ induced emotions such as shock, sadness, and especially guilt [P2, P3, P4, P7, P8]. Participants \textit{``felt worse''} [P8] after seeing their personal ChatGPT usage connected to environmental effects, and their usage patterns \textit{``really shocked [them]''} [P9]. Participants experienced heightened emotional responses for information and queries that could be accessed through Google or other means: \textit{``This extension made me feel a little guilty for turning to ChatGPT... I felt like the energy I could put into doing the research or work myself would be better spent rather than using ChatGPT.''} [P2]. Participants noted connections to current environmental concerns and personal sustainability goals: \textit{``Recent events like the LA wildfires has also made me more conscious of my environmental impact.''} [P5], \textit{``It made me realize that these things [ChatGPT] are not free, they must be paid for somewhere and this is paid in resources''} [P6]. In our design, these negative emotions are the catalyst for inducing behavioral change. 

Despite the negative emotions, participants enjoyed using \Sys, and rated it an average of 4.22 out of 5 (SD = 0.667, median = 4) on the Likert scale, which was an increase from  3.78 (SD = 1.084, median = 3.5) during pilot studies without visuals. With the improved user interface, fewer participants responded with themes of annoyance, and they felt more inclined to continue using the extension.

\subsection{Difficult to Overcome Utility of LLMs}

Four participants had a decrease in ChatGPT usage during the trial period [P2, P4, P7, P9], but the rest did not: one participant experienced no change [P3], and the remaining four participants actually increased their ChatGPT usage during the trial period [P1, P5, P6, P8]. In fact, the total queries across all participants increased by 18.584\% during the trial, though this is likely due to the small sample size and short trial period.

Many participants reported that the utility of ChatGPT outweighed their desire to reduce their environmental impact [P1, P3, P4, P5, P7]. Although it made them \textit{``more motivated to minimize [their] usage''} their \textit{``desperation and need will still make me continue using it''} [P5]. The two participants with the highest increased usage during the study  (21.429\% [P1] and 1,500\% [P5]) both acknowledged this increase in the exit-survey, citing outside factors like \textit{``apply[ing] to jobs''} [P1] and \textit{``necessary writing''} [P5].

Even when users did experience a decrease in ChatGPT use, they still valued the utility of ChatGPT: using \Sys\ was \textit{``mildly upsetting,''} but \textit{``Not upsetting enough to swear off ChatGPT altogether''} [P7]. Participants also cited a perceived limitation of individual actions to conserve resources, pointing out that their individual use is not a significant contributor to the overall impact of ChatGPT [P2, P4, P5], and expressing a desire for companies to take responsibility for the environmental impacts of LLMs, rather than users [P7]. One participant asked \textit{``how are we supposed to mitigate its energy consumption''} from an \textit{``existential''} perspective rather than a personal one [P4], though their usage did decline by 28.571\% over the course of the study. Taken together, these user reactions indicate of a key limitation of our intervention style. ChatGPT can be very useful, and appears to have no direct negative impact on its users. Our intervention relies on user awareness of the direct negative impacts ChatGPT can have on users, but the user makes the final decision on whether or not to continue using ChatGPT. For many participants, the utility of ChatGPT appeared to outweigh its detrimental effects on the environmental. However, in the week-long trial period, the total resource consumption lacked the chance to increase significantly---perhaps, over a longer period, the increased resource consumption would impact users more.

Interestingly, the limit popup did occasionally seem to affect behavior, whether or not participants realized. In total, popups appeared 10 times, and participants always closed it within a minute of its opening. For 7 of those popups, another query occurred less than 10 minutes after the popup opening. However, the other 3 popups led to a longer delay, ranging from 24 minutes to over 24 hours. This pattern was even more pronounced during pilot studies, when 21.429\% of popups preceded a 60+ minute query delay, and an additional 14.288\% preceded a 10+ minute delay. Thus, the popup may serve as a final push when users are already considering taking a break from ChatGPT. Allowing users to personalize their limits could increase this effect in future versions of \Sys.

\section{Discussion}

Contrary to our expectations, ChatGPT usage did not uniformly decrease during the study. One key reason we discovered in our analysis is that the utility of ChatGPT outweighs participants' desire to reduce personal environmental impacts. Eco-feedback systems tend to discourage negative behavior that has either minimal benefit or a negative impact on participants, like using more gas when driving a car \cite{EcoEffPref}. ChatGPT, in contrast, offers significant benefits, with little direct negative impact on the user, another limitation of metric-based systems \cite{ImproveEcoFeedback}. Additionally, participants felt desensitized to the information displayed and the reminders, a common issue in habit-reduction systems \cite{BeyondEnergyFeedback, GoalKeeper, LockoutPhones}. Although these issues limited our system's effectiveness in changing behavior, methods besides a limit popup and sidebar might improve results, and warrant future research. However, as participants indicate that they are inclined to keep using \Sys\ beyond the study period, it does increase user awareness even when behavior is dictated by external circumstances such as deadlines. 

\section{Limitations and Future Work}

We explicitly designed \Sys\ to have minimal privacy concerns, but the change in behavior it aims to elicit may have negative consequences for individuals, such as generalized skepticism and negative sentiments about emerging technology. Going forward, we could mitigate this potential concern by increasing user agency, focusing on flexible personal regulation of resource consumption, and avoiding criticism of emerging technology at large. Furthermore, increased awareness can place public pressure on companies to be more environmentally conscious, but \Sys\  could seem to blame individuals over large corporations. We believe that despite this risk, addressing the lack of public awareness about the environmental impact of LLMs is necessary to encourage meaningful change in the industry. 

We note that our sample consists entirely of university students who are regular users of ChatGPT, and are therefore inclined to query when faced with certain academic tasks and deadlines. In choosing active users, we wanted to ensure adequate testing of \Sys's features in the short study time. However, user behavior would likely be very different for those less familiar with ChatGPT, or those who use it for other tasks, such as personal or job-related matters. Furthermore, university students represent a highly educated subset of the population, but an ideal deployment of the system would occur at a broader scale. The effects of \Sys\ on other populations, who may have less pre-existing knowledge about LLMs or differing awareness of environmental threats, warrants further research.

Future work can continue exploring what factors encourage behavioral change. While some users with high pause times reported turning to alternative sources like \textit{``Google''} [P4] or \textit{``Google Scholar''} [P3] instead of ChatGPT, supporting our use of pop-up persistence time as a proxy for seeking alternate resources, others with similar pause patterns did not report platform-switching. Without better metrics on search product usage during pop-up periods, measuring its impact on behavioral change remains difficult. Although participants in the full study expressed far less annoyance than during pilot studies without visuals, they did request size-adjustment capabilities for the side panel, which sometimes blocked parts of the ChatGPT interface, even after moving it around the screen. Interestingly, one participant commented \textit{``sometimes [the side panel] can get in the way...unless of course that is the point''}, and another said \textit{``sometimes, the extension blocked parts of my screen, but this also made me lessen my usage of ChatGPT''} [P2]. These reflect research on input manipulation---decreasing a given behavior by making that behavior more difficult \cite{InputManipulation}. One method of input manipulation in \Sys\ could be modifying the size of the side panel. Participants also suggested incorporating animations and more images [P3, P7], as they thought visuals were the most easily understandable representation of their environmental impact. Additional visuals could also encourage \Sys\ users to compare their Eco Score and resource consumption with friends, facilitating a social comparison component that has been successful for other eco-feedback systems \cite{SocialSmartMeter}.

Future work must also consider a wider variety of LLMs, particularly new chain-of-thought models. Notably, these frontier models advertise more efficient training methods, but preliminary experiments suggest the interference cost per query is significant higher \cite{frontierModels}. Currently, per-query cost estimates remain unavailable for newer models, so determining these values is a key first step towards increasing consumer awareness.

\section{Conclusion}
The environmental impacts of LLMs remain largely unknown to their users. In an effort to increase user awareness and help mitigate the misuse of environmental resources in LLM use, we built \Sys, a novel Chrome extension that integrates into the ChatGPT website to provide live updates on personal environmental impact from querying. The extension provides live, anonymized tracking of user query activity and presents users with their individual environmental impact through dynamic, contextualized displays. We conducted a full study with nine participants over 7 days to evaluate our system. We found that the updating and contextualized statistics and visuals were valuable for user awareness. \Sys\ elicited strong emotional reactions from users, but often failed to cause behavior changes, due to the utility of ChatGPT. Already, our system demonstrates promising increases in user awareness, and we believe that further versions, with new features like social comparison and personalized goal setting, have high potential to improve user behavior and mitigate the environmental impacts of LLMs. 

\bibliographystyle{ACM-Reference-Format}
\bibliography{references}

\newpage

\appendix 
\onecolumn
\section{Tabular Data}
\label{apdx: tabular data}
\begin{table*}[h]
\begin{tabular}{ c|c|c }
 Participant & No. of Queries Per Week Before Trial & No. of Queries Per Week During Trial \\ 
 \hline \hline
 
 P1 & 42 & \cellcolor{red}51\\
 P2 & 34 & \cellcolor{green}29\\
 P3 & 7 & 7\\
 P4 & 14 & \cellcolor{green}10\\
 P5 & 1 & \cellcolor{red}16\\
 P6 & 0 & \cellcolor{red}5\\
 P7 & 7 & \cellcolor{green}6\\
 P8 & 0 & \cellcolor{red}3\\
 P9 & 8 & \cellcolor{green}7\\
 \end{tabular}
\caption{Summary of ChatGPT usage. Red cells represent more queries during the trial period than before, and green cells represent fewer queries during the trial than before.}
\label{tbl:usage appendix}
\end{table*}

\begin{table*}[h]
\begin{tabular}{ c|c|c|c|c|c|c|c }
 Participant & \thead[l]{Care for the \\ Environment \\ Pre Trial} & \thead[l]{Care for the \\ Environment \\ Post Trial} & \thead[l]{Care about the \\ Environmental \\ Impact of Queries \\ Pre Trial} & \thead[l]{Care about the \\ Environmental \\ Impact of Queries \\ Post Trial} & Enjoyment & \thead[l]{Likelihood of \\ Future Use} & \thead[l]{Learned \\ More}\\
  \hline \hline
 
 P1 & 4&5&3&3&4&4&3\\
 P2 & 4&4&2&3&4&3&4\\
 P3 & 5&4&2&4&4&3&4\\
 P4 & 4&4&4&4&5&5&4\\
 P5 & 4&4&3&4&4&5&4\\
 P6 & 3&3&2&2&5&4&5\\
 P7 & 5&5&5&5&5&3&5\\
 P8 & 3&4&3&4&3&1&5\\
 P9 & 4&4&3&3&4&3&3\\
 \end{tabular}
\caption{Responses to multiple choice survey questions on a Likert scale of 1 to 5.}
\label{tbl:survey responses}
\end{table*}

\begin{table*}[h]
\begin{tabular}{ c|c|c }
 Participant & Gender & Age\\
  \hline \hline
 
 P1 & Female&18 \\
 P2 & Female& 21\\
 P3 & Non Binary&20\\
 P4 &Female&19 \\
 P5 & Female&21\\
 P6 & Male&24\\
 P7 & Female&21\\
 P8 & Male&20\\
 P9 & Female&21\\
 \end{tabular}
\caption{Demographic data for participants.}
\label{tbl:demographics}
\end{table*}

\clearpage
\section{Eco Score Algorithm}

\begin{algorithm}
\label{asec:alg}
\caption{Eco Score Logic for Queries. This algorithm runs each time a new query is detected. In addition to this algorithm, Eco Score will also automatically increase by 1 point every 20 minutes.}\label{alg:eco_score}
\begin{algorithmic}
\STATE $pauseLength \gets currentQueryTime - previousQueryTime$
\IF{$pauseLength \geq 60$}
    \STATE $ecoScore \gets ecoScore - 7$
\ELSIF{$pauseLength \geq 30$}
    \STATE $ecoScore \gets ecoScore - 8$
\ELSIF{$pauseLength \geq 15$}
    \STATE $ecoScore \gets ecoScore - 9$
\ELSIF{$pauseLength \geq 7$}
    \STATE $ecoScore \gets ecoScore - 10$
\ELSIF{$pauseLength \geq 3$}
    \STATE $ecoScore \gets ecoScore - 11$
\ELSIF{$pauseLength \geq 1$}
    \STATE $ecoScore \gets ecoScore - 12$
\ELSE
    \STATE $ecoScore \gets ecoScore - 13$
\ENDIF
\IF{$ecoScore < 0$}
    \STATE $ecoScore \gets 0$
\ENDIF
\end{algorithmic}
\end{algorithm}

\section{Codebook During Qualitative Analysis}
\label{apdx: codebook}
\includegraphics[width=0.75\linewidth]{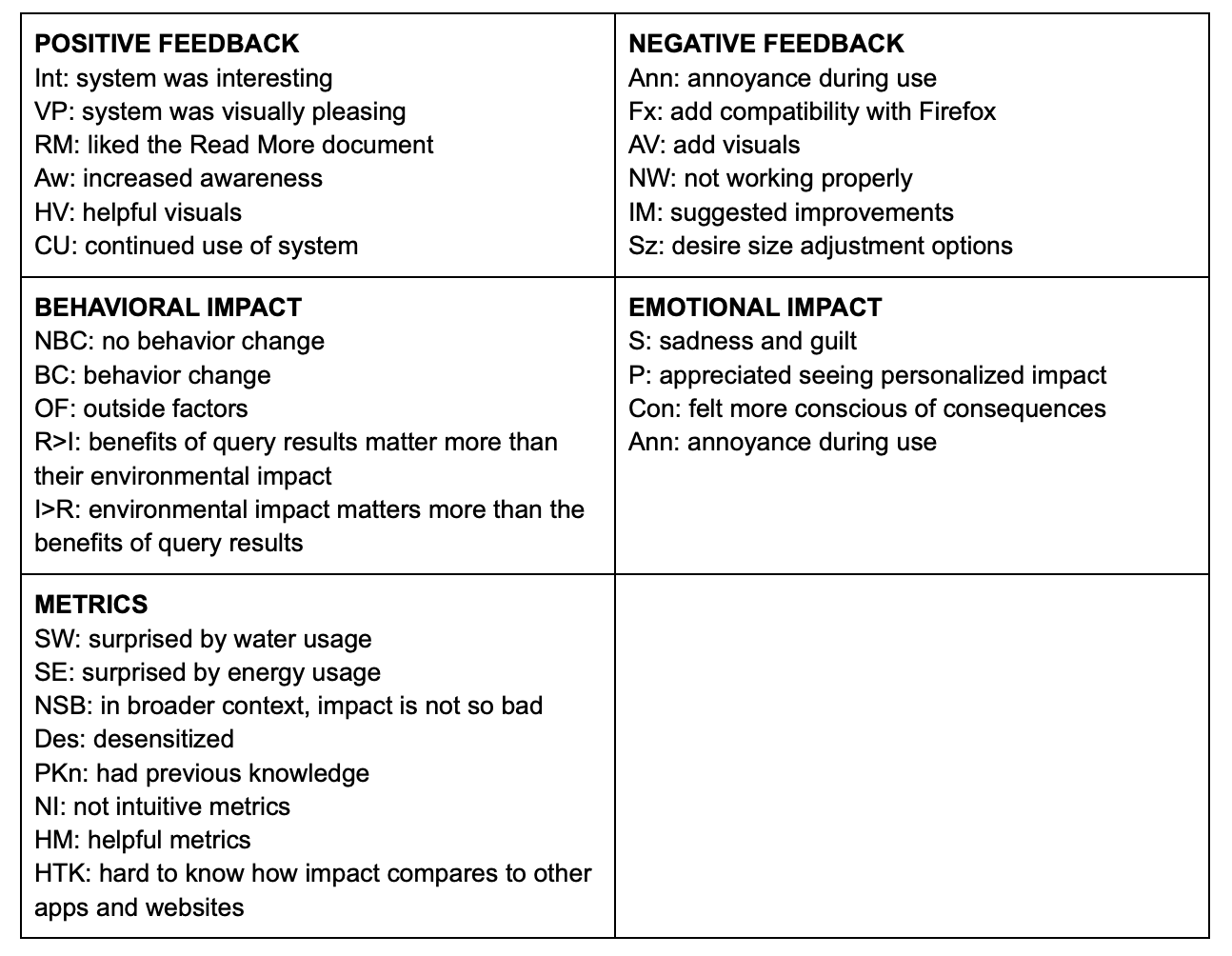}
\Description[Image of all codes and their meanings, as used by researchers during qualitative analysis coding.]{Section 1: Positive Feedback. Int: system was interesting. VP: system was visually pleasing. RM: liked the Read More document. Aw: increased awareness. HV: helpful visuals. CU: continued use of system. Section 2: Negative Feedback. Ann: annoyance during use. Fx: add compatibility with Firefox. AV: add visuals. NW: not working properly. IM: suggested improvements. Sz: desire size adjustment options. Section 3: Behavioral Impact. NBC: no behavior change. BC: behavior change. OF: outside factors. R>I: benefits of query results matter more than their environmental impact. I>R: environmental impact matters more than the benefits of query results. Section 4: Emotional Impact. S: sadness and guilt. P: appreciated seeing personalized impact. Con: felt more conscious of consequences. Ann: annoyance during use. Section 5: Metrics. SW: surprised by water usage. SE: surprised by energy usage . NSB: in broader context, impact is not so bad. Des: desensitized. PKn: had previous knowledge. NI: not intuitive metrics. HM: helpful metrics. HTK: hard to know how impact compares to other apps and websites.}

\clearpage
\section{Surveys}
\subsection{Pre-Survey}
\label{apdx: pre survey}
\begin{figure}[h]
    \centering
    \begin{subfigure}[t]{0.49\textwidth}
        \centering
        \includegraphics[width=\textwidth]{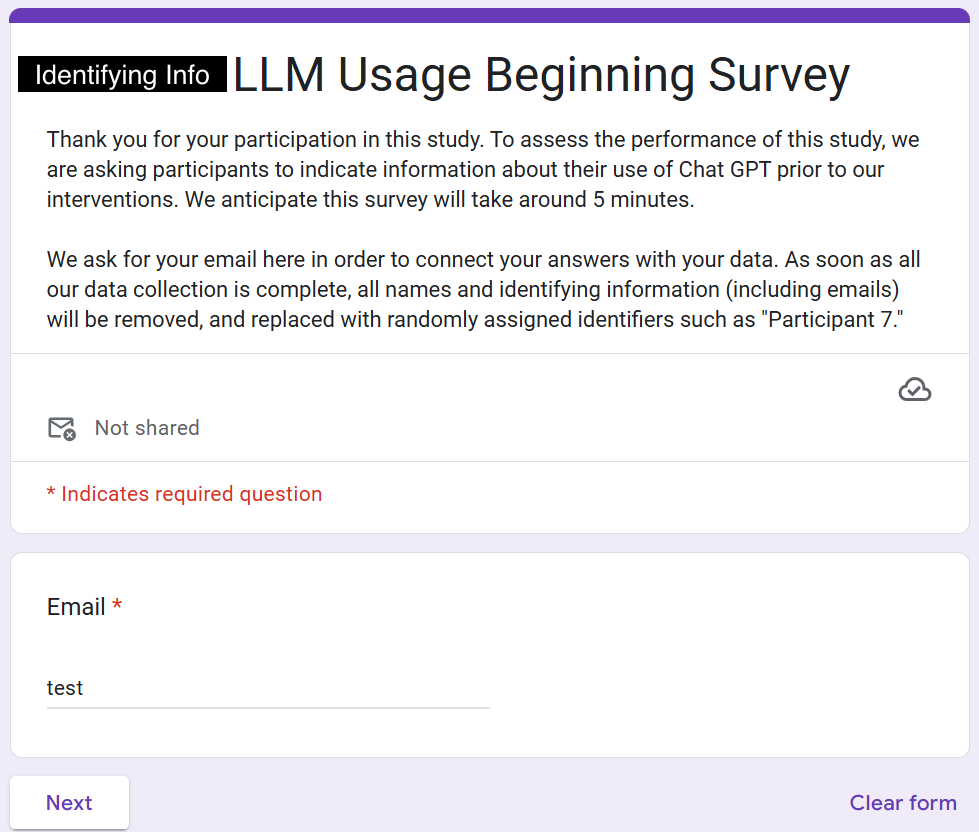}
        \caption{Pre-Survey Page 1}
        \Description[Page 1 of Pre-Survey]{Form title: LLM Usage Beginning Survey. Introduction text: Thank you for your participation in this study. To assess the performance of this study, we are asking participants to indicate information about their use of Chat GPT prior to our interventions. We anticipate this survey will take around 5 minutes. We ask for your email here in order to connect your answers with your data. As soon as all our data collection is complete, all names and identifying information (including emails) will be removed, and replaced with randomly assigned identifiers such as ``Participant 7.'' Question (required): Email.}
    \end{subfigure}
    \hfill
    \begin{subfigure}[t]{0.49\textwidth}
        \centering
        \includegraphics[width=\textwidth]{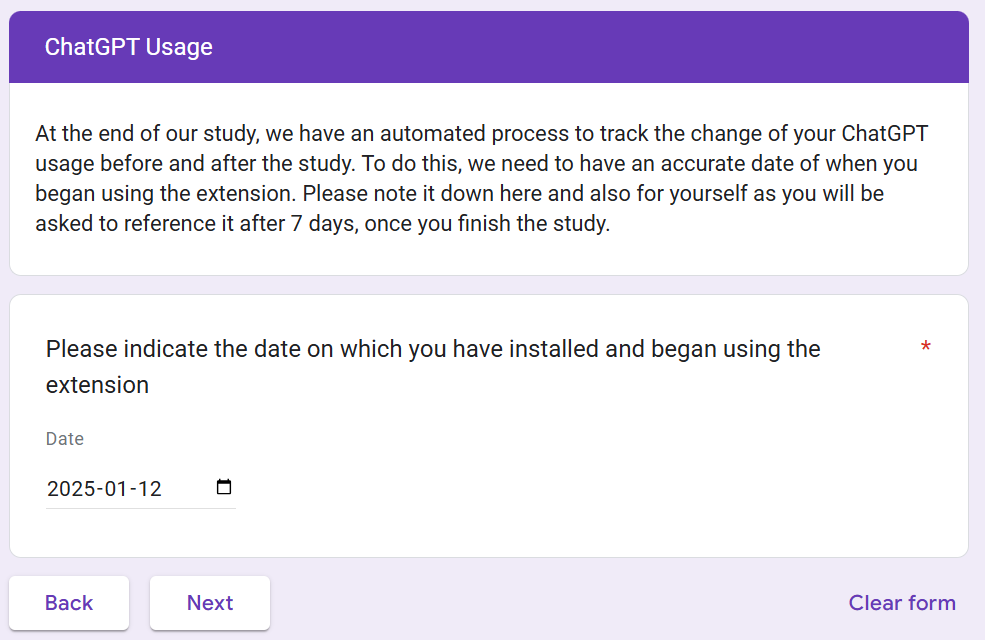}
        \caption{Pre-Survey Page 2}
        \Description[Page 2 of Pre-Survey]{Section title: ChatGPT Usage. Introduction text: At the end of our study, we have an automated process to track the change of your ChatGPT usage before and after the study. To do this, we need to have an accurate date of when you began using the extension. Please note it down here and also for yourself as you will be asked to reference it after 7 days, once you finish the study. Question (required): Please indicate the date on which you have installed and began using the extension.}
    \end{subfigure}
    \begin{subfigure}[t]{0.49\textwidth}
        \centering
        \includegraphics[width=\textwidth]{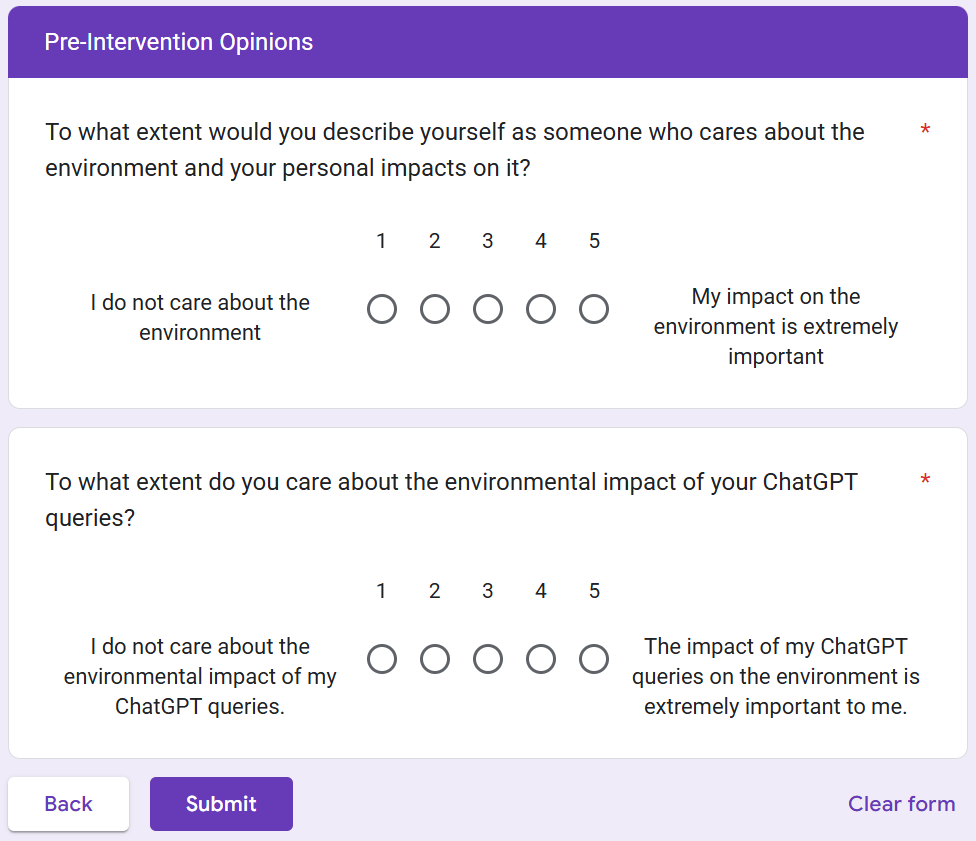}
        \caption{Pre-Survey Page 3}
        \Description[Page 3 of Pre-Survey]{Section title: Pre-Intervention Opinions. Question (required): To what extent would you describe yourself as someone who cares about the environment and your personal impacts on it? This question provides multiple choice options ranging from 1 (I do not care about the environment) to 5 (My impact on the environment is extremely important). Question: (required): To what extent do you care about the environmental impact of your ChatGPT queries? This question provides multiple choice options ranging from 1 (I do not care about the environmental impact of my ChatGPT queries) to 5 (The impact of my ChatGPT queries on the environment is extremely important to me).}
    \end{subfigure}
    \caption{Study Participant Pre-Survey}
\end{figure}
\newpage
\subsection{Post-Survey}
\label{apdx: post survey}
\begin{figure}[h]
    \centering

    \begin{minipage}[t]{0.46\textwidth}
        \centering
        \includegraphics[width=\textwidth]{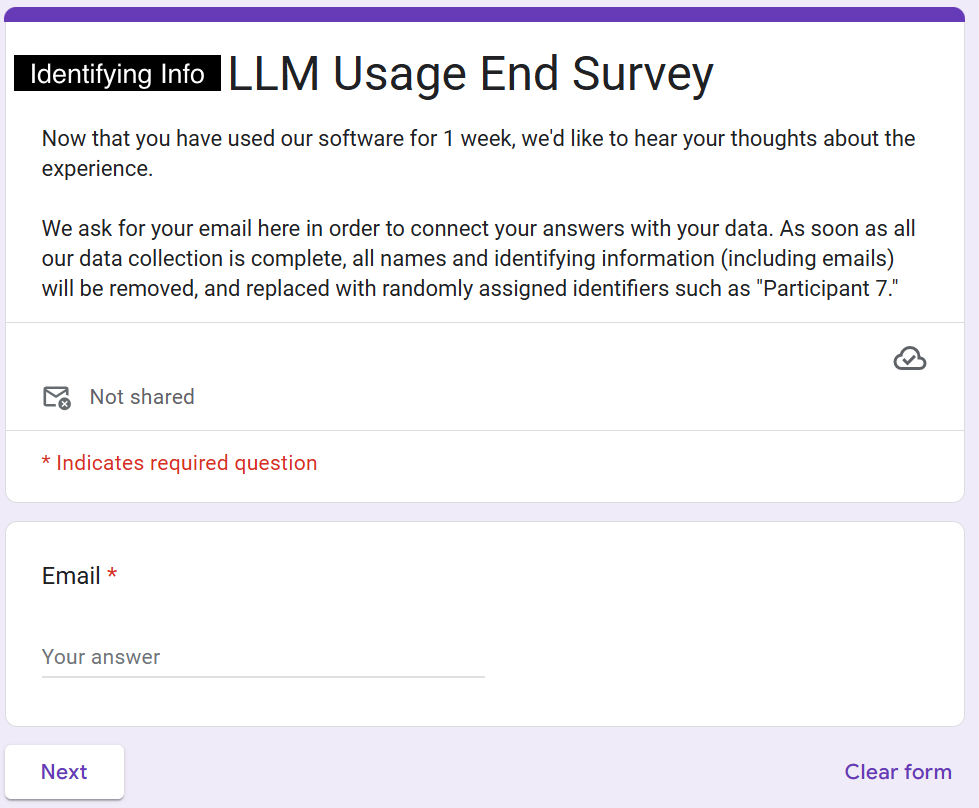}
        \caption*{Post-Survey Page 1}
        \Description[Page 1 of Post-Survey]{Form title: LLM Usage End Survey. Introduction text: Now that you have used our software for 1 week, we'd like to hear your thoughts about the experience. We ask for your email here in order to connect your answers with your data. As soon as all our data collection is complete, all names and identifying information (including emails) will be removed, and replaced with randomly assigned identifiers such as ``Participant 7.'' Question (required): Email.}
    \end{minipage}
    \hfill
    \begin{minipage}[t]{0.46\textwidth}
        \centering
        \includegraphics[width=\textwidth]{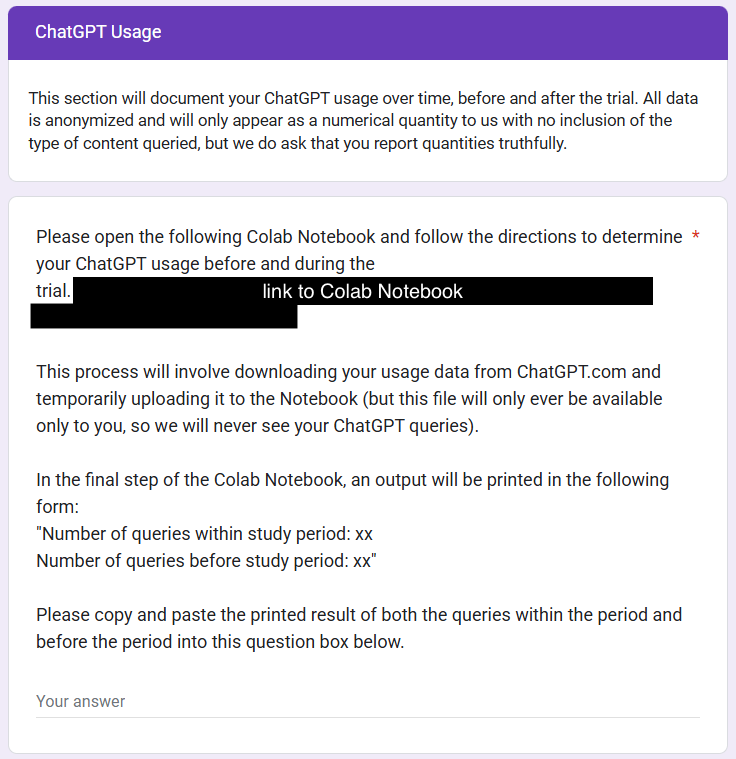}
        \caption*{Post-Survey Page 2}
        \Description[Page 2 of Post-Survey]{Section title: ChatGPT Usage. Introduction text: This section will document your ChatGPT usage over time, before and after the trial. All data is anonymized and will only appear as a numerical quantity to us with no inclusion of the type of content queried, but we do ask that you report quantities truthfully. 
        Question (required): Please open the following Colab Notebook and follow the directions to determine your ChatGPT usage before and during the trial (link). This process will involve downloading your usage data from ChatGPT.com and temporarily uploading it to the Notebook (but this file will only ever be available only to you, so we will never see your ChatGPT queries). In the final step of the Colab Notebook, an output will be printed in the following form: ``Number of queries within study period: xx \n Number of queries before study period: xx'' Please copy and paste the printed result of both the queries within the period and before the period into this question box below. }
    \end{minipage}

    \vspace{1em} 
    \begin{minipage}[t]{0.46\textwidth}
        \centering
        \includegraphics[width=\textwidth]{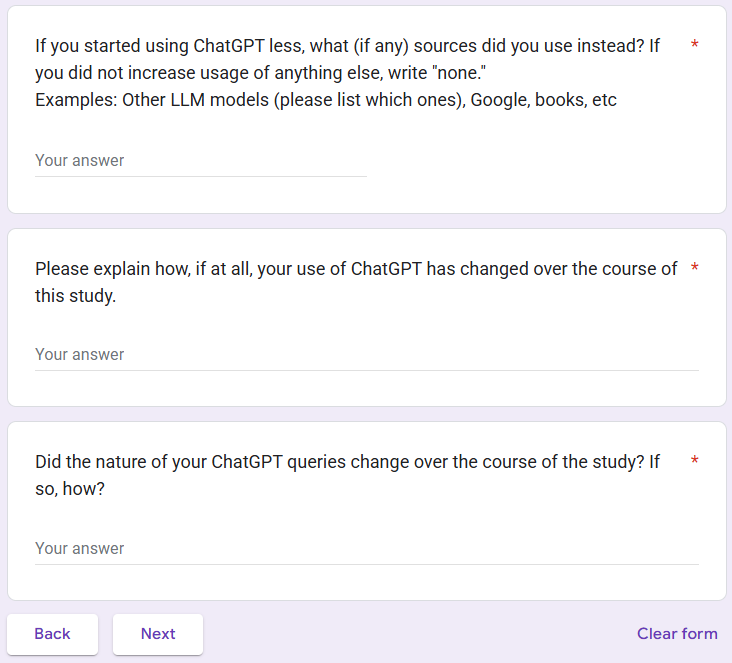}
        \caption*{Post-Survey Page 3}
        \Description[Page 3 of Post-Survey]{Question (required): If you started using ChatGPT less, what (if any) sources did you use instead? If you did not increase usage of anything else, write "none." Examples: Other LLM models (please list which ones), Google, books, etc. Question (required): Please explain how, if at all, your use of ChatGPT has changed over the course of this study. Question (required): Did the nature of your ChatGPT queries change over the course of the study? If so, how?}
    \end{minipage}
    \hfill
    \begin{minipage}[t]{0.46\textwidth}
        \centering
        \includegraphics[width=\textwidth]{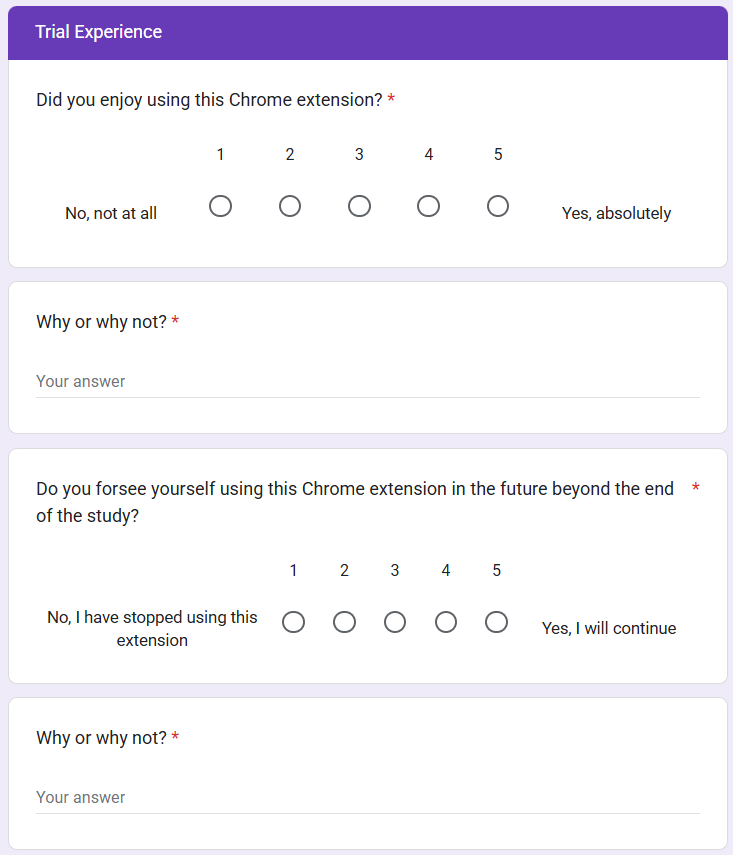}
        \caption*{Post-Survey Page 4}
        \Description[Page 4 of Post-Survey]{Section title: Trial Experience. Question (required): Did you enjoy using this Chrome extension? Options range from 1 (No, not at all) to 5 (Yes, absolutely). Question (required): Why or why not? Question (required): Do you forsee yourself using this Chrome extension in the future beyond the end of the study? Options range from 1 (No, I have stopped using this extension) to 5 (Yes, I will continue). Question (required): Why or why not?}
    \end{minipage}

    \caption{Study Participant Post-Survey}
\end{figure}

\clearpage 

\begin{figure}[h]
    \centering

    \begin{minipage}[t]{0.47\textwidth}
        \centering
        \includegraphics[width=\textwidth]{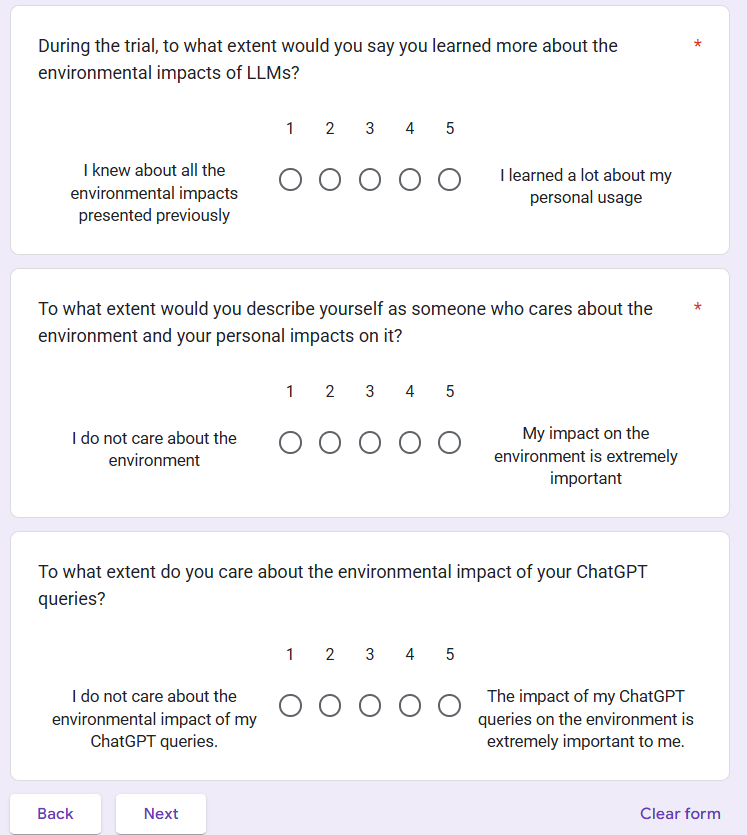}
        \caption*{Post-Survey Page 5}
        \Description[Page 5 of Post-Survey]{Question (required): During the trial, to what extent would you say you learned more about the environmental impacts of LLMs? Options range from 1 (I knew about all the environmental impacts presented previously) to 5 (I learned a lot about my personal usage). Question (required): To what extent would you describe yourself as someone who cares about the environment and your personal impacts on it? Options range from 1 (I do not care about the environment) to 5 (My impact on the environment is extremely important). Question (required): To what extent do you care about the environmental impact of your ChatGPT queries? Options range from 1 (I do not care about the environmental impact of my ChatGPT queries) to 5 (The impact of my ChatGPT queries on the environment is extremely important to me).}
    \end{minipage}
    \hfill
    \begin{minipage}[t]{0.47\textwidth}
        \centering
        \includegraphics[width=\textwidth]{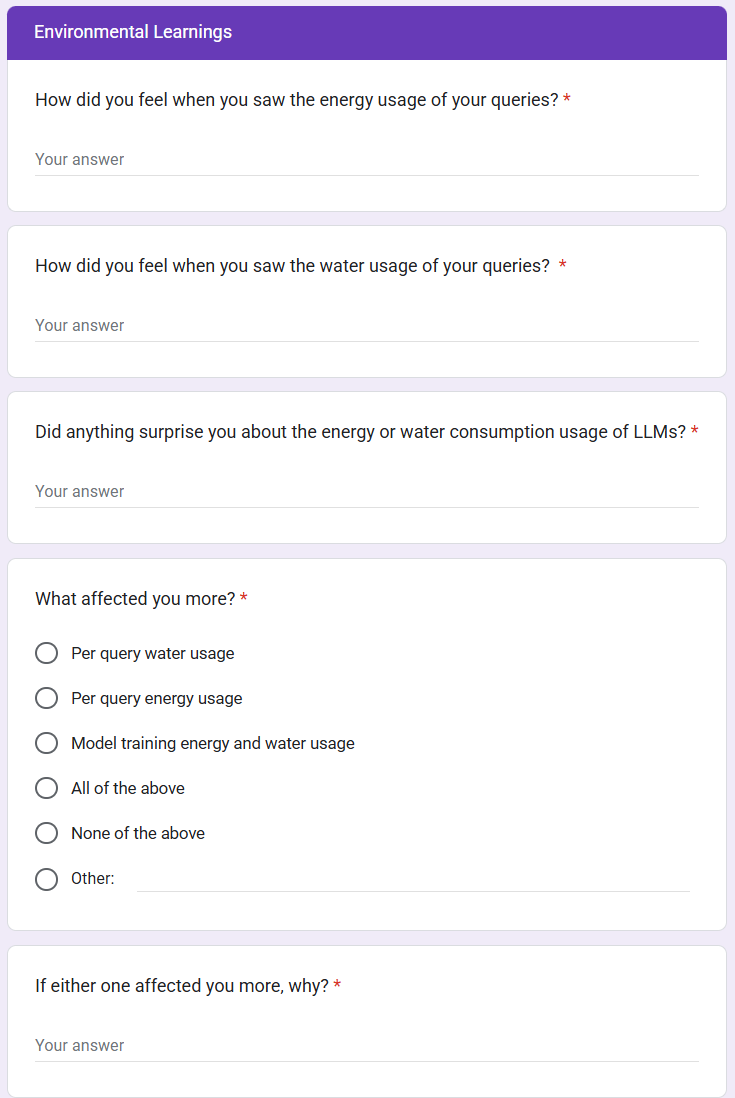}
        \caption*{Post-Survey Page 6}
        \Description[Page 6 of Post-Survey]{Section title: Environmental Learnings. Question (required): How did you feel when you saw the energy usage of your queries? Question (required): How did you feel when you saw the water usage of your queries? Question (required): Did anything surprise you about the energy or water consumption usage of LLMs? Question (required): What affected you more? Multiple choice options: per query water usage, per query energy usage, model training energy and water usage, all of the above, none of the above, other (with space to input a custon response). Question (required): If either one affected you more, why?}
    \end{minipage}

    \vspace{1em} 
    \begin{minipage}[t]{0.47\textwidth}
        \centering
        \includegraphics[width=\textwidth]{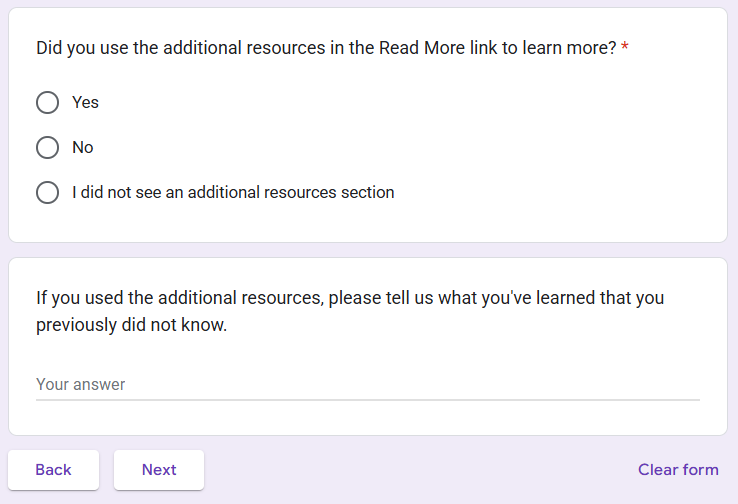}
        \caption*{Post-Survey Page 7}
        \Description[Page 7 of Post-Survey]{Question (required): Did you use the additional resources in the Read More link to learn more? Multiple choice options: yes, no, I did not see an additional resources section. Question (not required): If you used the additional resources, please tell us what you've learned that you previously did not know.}
    \end{minipage}
    \hfill
    \begin{minipage}[t]{0.47\textwidth}
        \centering
        \includegraphics[width=\textwidth]{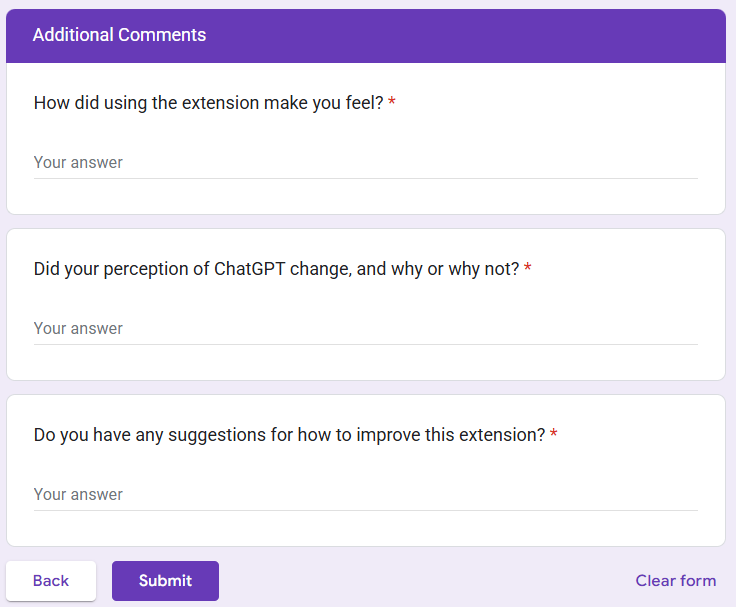}
        \caption*{Post-Survey Page 8}
        \Description[Page 8 of Post-Survey]{Section title: Additional Comments. Question (required): How did using the extension make you feel? Question (required): Did your perception of ChatGPT change, and why or why not? Question (required): Do you have any suggestions for how to improve this extension?}
    \end{minipage}

    \caption{Study Participant Post-Survey (cont.)}
\end{figure}

\clearpage 

\section{Query Counting Colab Notebook}
\label{apdx: colab notebook}
Included in the post-trial survey, users were prompted to upload their conversation data and run this notebook to count their queries. Participants are only asked to report the final numerical counts so no conversation data is ever seen by anyone other than the user. Following are the instructions participants received:

In this step, you will upload your personal ChatGPT history to this notebook. It will be saved locally and temporarily, and no one else will ever have access to it.
\begin{enumerate}
    \item Open ChatGPT.com and navigate to Settings by clicking on your profile image.
    \item Select Data Controls on the left sidebar of Settings, then click Export Data. This will send an export link to your email. Click that link, then unzip the downloaded folder.
    \item In this Google Colab Notebook, open Files on the left sidebar. Click the Upload button, and upload the file named \texttt{conversations.json} from the folder you just exported.
    \item Run the following code block. If there is an error, check that you successfully uploaded \texttt{conversations.json}. If you need any further assistance, contact one of the experimenters.
\end{enumerate}
\subsection*{Checking Uploaded Data}
\begin{verbatim}
import pandas as pd
import json

# check that the data was uploaded successfully
# this code block should run without errors
df = pd.read_json('conversations.json')
data = df.to_dict(orient='records')
\end{verbatim}

\subsection*{Calculate Weekly Queries}
Input the date you downloaded the Chrome Extension into the following code block, then run it.
\subsection*{Setting the Download Date}
\begin{verbatim}
download_date_str = '2025-01-19' # @param {type:"date"}
\end{verbatim}
\subsection*{Analyzing Query Counts}
\begin{verbatim}
import datetime

# convert date string to universal time
download_date = download_date_str.split('-')
download_date = datetime.datetime(int(download_date[0]), int(download_date[1]), int(download_date[2]),
    tzinfo=datetime.timezone.utc)

from datetime import timedelta

# global variables
study_query_count = 0
historical_query_count = 0
pre_trial = download_date - timedelta(days=7)
during_trial = download_date + timedelta(days=7)

# loop through json
for conversation in data:
    for message_id, message_details in conversation.get('mapping', {}).items():
        # Check if the message is authored by the user
        message = message_details.get('message')
        if message and message.get('author', {}).get('role') == 'user':
            query_time = message.get('create_time')
            # convert query_time from float to datetime
            query_time = datetime.datetime.fromtimestamp(query_time, tz=datetime.timezone.utc)
            # check which time frame query was in
            if pre_trial <= query_time < download_date:
              historical_query_count += 1
            elif download_date <= query_time <= during_trial:
              study_query_count += 1

print(f"Number of queries within study period: {study_query_count}")
print(f"Number of queries before study period: {historical_query_count}")
\end{verbatim}

\section{Side Panel Graphics}
The picture on the side panel cycles through five different images, reflective of the Eco Score. For each 20 point bracket of the Eco Score (100-80, 79-60, etc.) the image will change to reflect a deteriorating environment. 
\begin{figure}[h]
    \centering
    \begin{subfigure}[t]{0.19\textwidth}
        \centering
        \includegraphics[width=\textwidth]{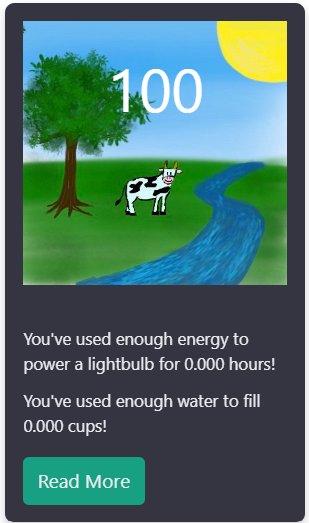}
        \caption{Side Panel Image 1}
        \Description[Eco Score of 100]{This image shows an illustrated image of a cow, a flowing stream, a green tree, and a bright blue sky, overlaid with an example EcoScore of 100. The text reads ``You've used enough energy to power a lightbulb for 0.000 hours! You've used enough water to fill 0.000 cups!''. At the bottom is a green button labeled ``Read More.''}
    \end{subfigure}
    \begin{subfigure}[t]{0.19\textwidth}
        \centering
        \includegraphics[width=\textwidth]{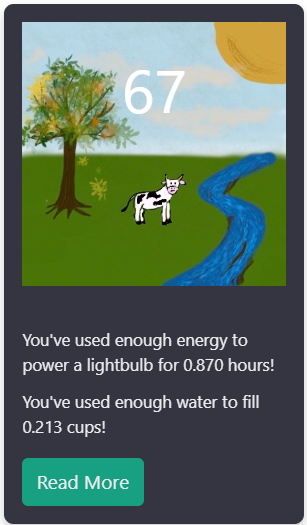}
        \caption{Side Panel Image 2}
        \Description[Eco Score of 67]{This image shows an illustrated image of a skinny cow, a stream with some debris, a green and yellow tree, and a blue sky with faint clouds of pollution, overlaid with an example EcoScore of 67. The text reads ``You've used enough energy to power a lightbulb for 0.870 hours! You've used enough water to fill 0.213 cups!''. At the bottom is a green button labeled ``Read More.''}
    \end{subfigure}
    \begin{subfigure}[t]{0.19\textwidth}
        \centering
        \includegraphics[width=\textwidth]{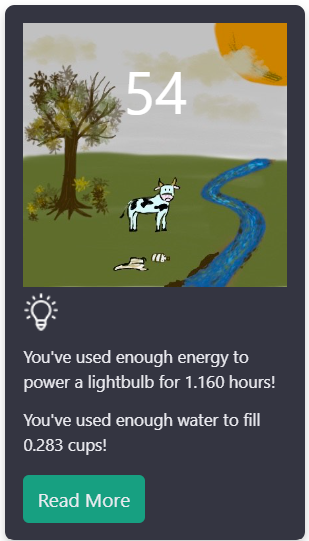}
        \caption{Side Panel Image 3}
        \Description[Eco Score of 54]{This image shows an illustrated image of a skinny cow next to some trash, a stream with lots of debris, a green, brown, and yellow tree, and a gray-blue sky with many brown clouds, overlaid with an example EcoScore of 54. The text reads ``You've used enough energy to power a lightbulb for 1.160 hours! You've used enough water to fill 0.283 cups!''. There is a pictogram of a single lightbulb. At the bottom is a green button labeled ``Read More.''}
    \end{subfigure}
    \begin{subfigure}[t]{0.19\textwidth}
        \centering
        \includegraphics[width=\textwidth]{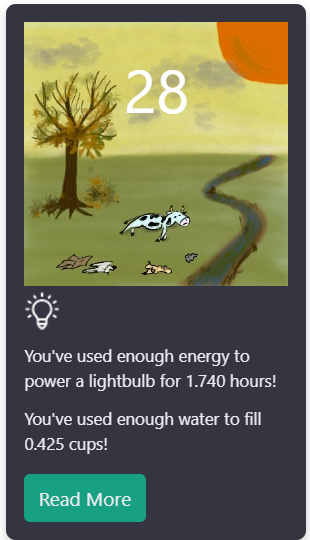}
        \caption{Side Panel Image 4}
        \Description[Eco Score of 28]{This image shows an illustrated image of a skinny cow lying on the ground with some trash, an almost completely dried-up stream, a brown and yellow tree, and yellow sky with clouds of pollution, overlaid with an example EcoScore of 28. The text reads ``You've used enough energy to power a lightbulb for 1.740 hours! You've used enough water to fill 0.425 cups!''. There is a pictogram of a single lightbulb. At the bottom is a green button labeled ``Read More.''}
    \end{subfigure}
    \begin{subfigure}[t]{0.19\textwidth}
        \centering
        \includegraphics[width=\textwidth]{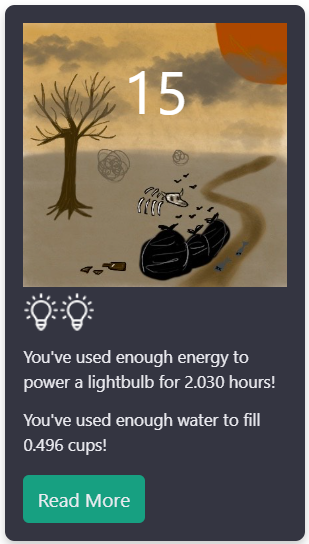}
        \caption{Side Panel Image 5}
        \Description[Eco Score of 15]{This image shows an illustrated image of a cow carcass with lots of trash, a dried-up stream, a dead tree, and yellow sky with clouds of pollution, overlaid with an example EcoScore of 15. The text reads ``You've used enough energy to power a lightbulb for 2.030 hours! You've used enough water to fill 0.496 cups!''. There is a pictogram of two lightbulbs. At the bottom is a green button labeled ``Read More.''}
    \end{subfigure}
\end{figure}
\section{Installation Instructions}
Participants were prompted to install their custom extensions linked with their user ID for the study from a ZIP file. 

\includegraphics[width=.8\textwidth]{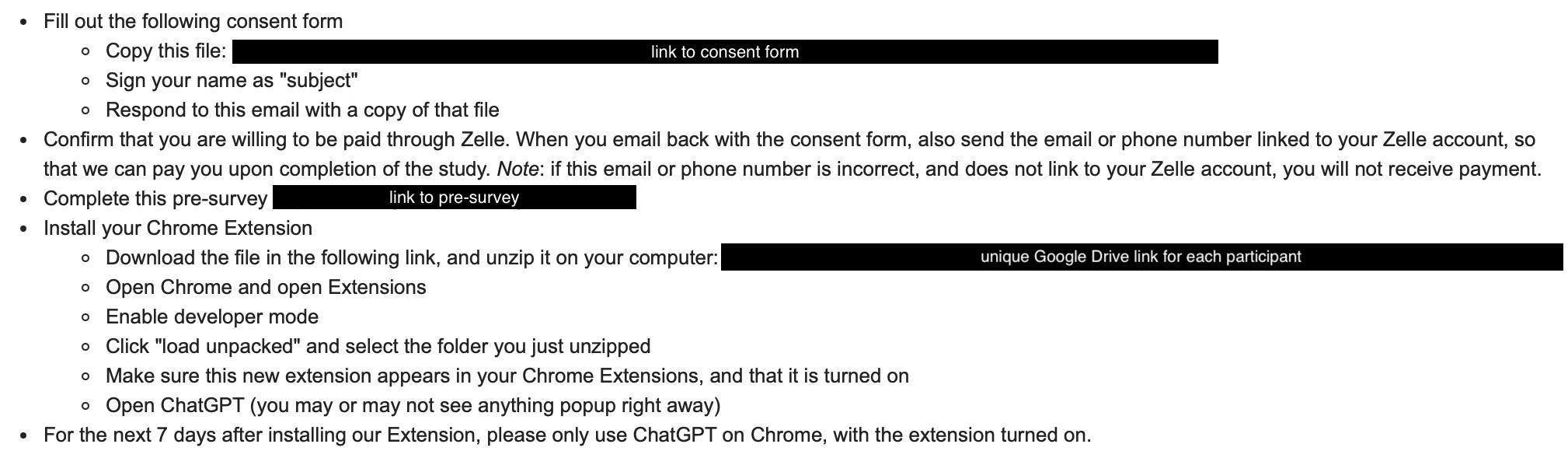}
\Description[Email instructions sent to participants]{This screenshot contains the following text: 1. Fill out the following consent form. 1a. Copy this file: (link). 1b. Sign your name as ``subject''. 1c. Respond to this email with a copy of that file. 2. Confirm that you are willing to be paid through Zelle. When you email back with the consent form, also send the email or phone number linked to your Zelle account, so that we can pay you upon completion of the study. Note: if this email or phone number is incorrect, and does not link to your Zelle account, you will not receive payment. 3. Complete this pre-survey (link). 4. Install your Chrome Extension. 4a. Download the file in the following link, and unzip it on your computer: (link). 4b. Open Chrome and open Extensions. 4c. Enable developer mode. 4d. Click "load unpacked" and select the folder you just unzipped. 4e. Make sure this new extension appears in your Chrome Extensions, and that it is turned on. 4f. Open ChatGPT (you may or may not see anything popup right away). 5. For the next 7 days after installing our Extension, please only use ChatGPT on Chrome, with the extension turned on.}

\section{Read More Document}
The Read More button in \Sys\  links to the following document of additional information.
\\

\includegraphics[width=.8\textwidth]{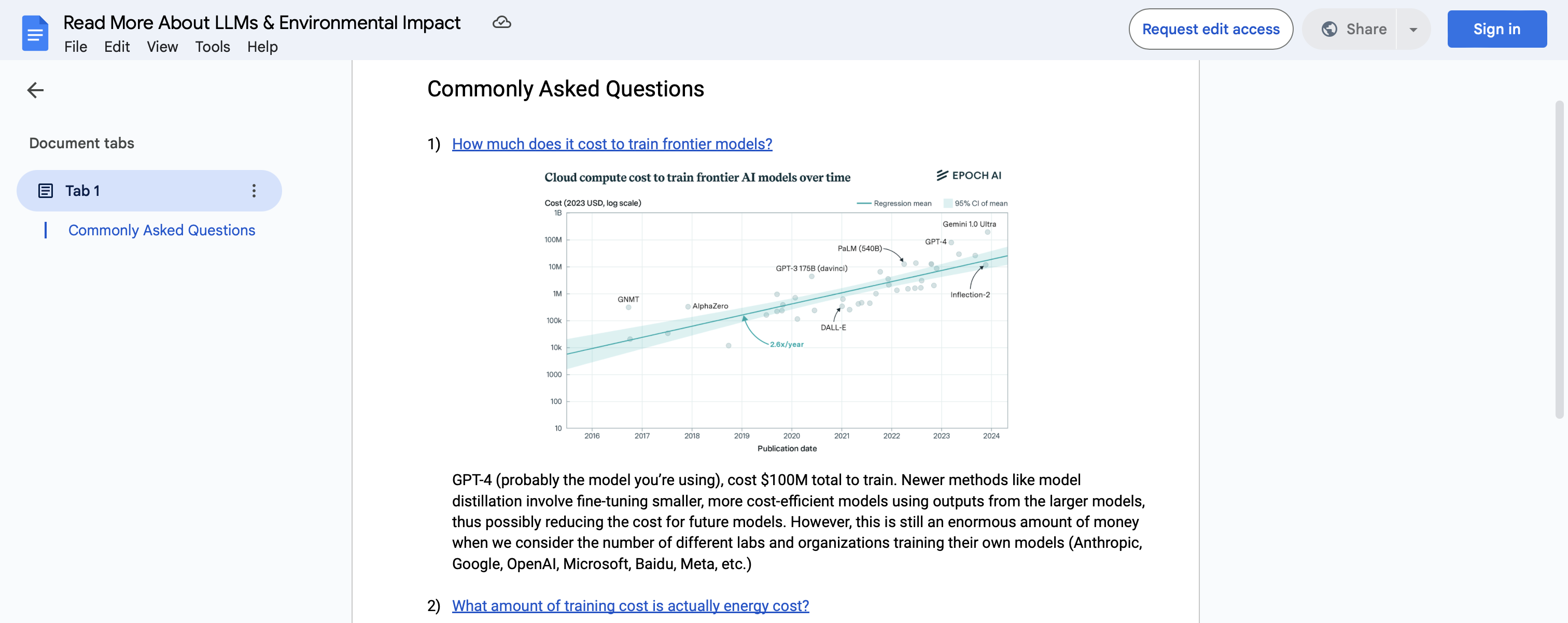}
\Description[Screenshot of Read More Document]{This partial screenshot of the Read More document shows a public Google Docs of Commonly asked Questions, including ``How much does it cost to train frontier models?'' and ``What amount of training cost is actually energy cost?'' There are links to sources, as well as brief answers with graphs.}

\section{Server Participant Tracking}
The following is a segment of the data logged by the extension during user studies. The full spreadsheet is private, and accessible only to researchers.

\includegraphics[width=0.5\linewidth]{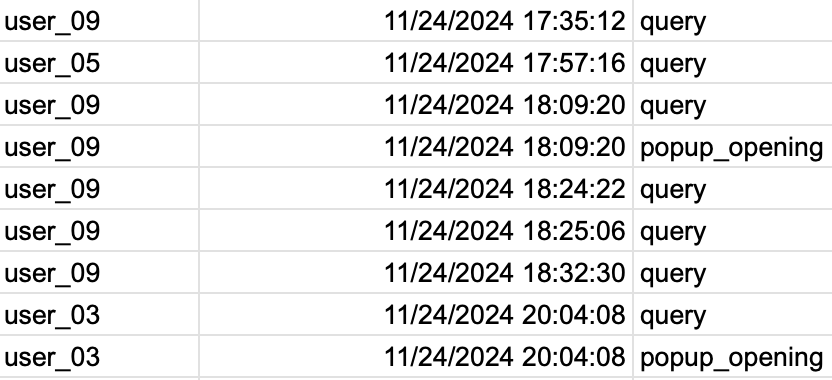}
\Description[Screenshot of spreadsheet of logged queries and popups]{This is a sample screenshot of the log of usage available to researchers during the trial period. One column lists participant IDs (i.e. user_09, user_05, etc). The second column lists the date and time of the logged event. The third column lists the label of the logged event (i.e. query, popup_opening, popup_closed, or readmore_clicked).}

\end{document}